\documentclass[preprint,aps]{revtex4}
\usepackage{graphicx}
\newcommand{\be}{\begin{eqnarray}}
\newcommand{\ee}{\end{eqnarray}}
\newcommand{\ket}[1]{\vert\,{#1}\rangle}
\newcommand{\eps}{\epsilon}
\begin{document}
\begin{flushright}
SLAC-PUB-11615, UFIFT-HEP-06-02, UCRL-JRNL-220485
\end{flushright}
%\vspace*{1cm}
\title
%{Proton Optics:  Light-front Wave Functions \\and Deeply Virtual
%Compton Scattering in Longitudinal Coordinate Space}
{Hadron Optics: \\ Diffraction Patterns in Deeply Virtual Compton Scattering}
\author{\bf S. J. Brodsky$^a$, D. Chakrabarti$^b$, A. Harindranath$^c$, A.
Mukherjee$^d$, J. P. Vary$^{e,a,f}$}
%\email{dipankar@phys.ufl.edu}
\affiliation{$^a$ Stanford Linear Accelerator Center, Stanford
University, Stanford, California 94309, USA.\\
$^b$ Department of Physics, University of Florida, Gainesville,
FL-32611-8440, USA.\\
$^c$ Saha Institute of Nuclear Physics, 1/AF Bidhannagar, Kolkata
700064, India. \\
$^d$ Department of Physics,
Indian Institute of Technology, Powai, Mumbai 400076,
India.\\
$^e$ Department of Physics and Astronomy,
Iowa State University, Ames, Iowa 50011, USA.\\
$^f$ Lawrence Livermore National Laboratory, \\
L-414, 7000 East Avenue,
Livermore, California, 94551, USA.}
%\date{\today\\[2cm]}
%\date{\today}
\date{31$^{\rm st}$ July, 2006}

\begin{abstract}
We show that the Fourier transform of the Deeply Virtual Compton
Scattering (DVCS) amplitude with respect to the skewness variable $\zeta$ 
at fixed invariant momentum transfer squared $t$
provides a unique way to visualize the structure of the target hadron
in the boost-invariant longitudinal coordinate space. The results are 
analogous to the diffractive scattering of a wave in optics. As a specific 
example, we utilize the quantum fluctuations of a fermion state at one loop 
in QED to obtain the behavior of the DVCS amplitude for electron-photon 
scattering. We then simulate the wavefunctions for a hadron by differentiating 
the above LFWFs with respect to $M^2$ and study the corresponding DVCS 
amplitudes in light-front longitudinal space. In both cases we observe that 
the diffractive patterns in the longitudinal variable conjugate to 
$\zeta$  sharpen and the positions of the first 
minima move in with increasing momentum transfer. For fixed $t$, 
higher minima  appear at positions which are integral multiples of the lowest 
minimum. Both these observations strongly support the  analogy with diffraction 
in optics. 
\end{abstract}
\maketitle

%%%%%%%%%%%%%%%%%%%%%%%%%%%%%%%%%%%%%%%%%%%%%%%%%%%%%%%%%%%%%%%%%%%%%%%%%%%
\noindent{\bf Introduction}
\vskip .1in
Deeply Virtual Compton Scattering (DVCS) $\gamma^*(q)~ +~p(P) \to
\gamma(q')~+~ 
p(P^\prime)$
provides a remarkable tool for studying the fundamental structure of
the proton at the amplitude level. We define the momentum transfer 
$\Delta=P-P'$ and the invariant momentum transfer squared 
$t=\Delta^\mu \Delta_\mu$. When the incoming photon is highly 
virtual $Q^2 = -q^2 >> \Lambda^2_{QCD}$ , the underlying scattering process 
measures Compton scattering on bound quarks, convoluted with the fundamental 
microscopic wavefunctions of the
initial- and final-state proton.  In addition, the initial-state photon can 
scatter on  virtual $q \bar q$  pairs in the target which are then 
annihilated by the final-state photon, thus probing the particle-number 
quantum fluctuations of the hadron wavefunction required for Lorentz 
invariance.   Measurements of the DVCS cross sections with specific proton 
and photon polarizations can provide comprehensive probes of the spin as 
well as spatial structure of the proton at the most fundamental level of QCD.

The theoretical analysis of DVCS is particularly clear and compelling
when one utilizes light-front quantization at fixed
$\tau= y^+$. (We use the standard LF
coordinates  $P^\pm = P^0 \pm P^3, y^\pm = y^0 \pm y^3$.   Since the proton
is on-shell, $P^+ P^--P^2_\perp = M^2_p ) .$ If we neglect radiative corrections to the struck quark
propagator (i.e., set the Wilson line to $1$), then the required DVCS quark
matrix elements can  be computed from the overlap of the boost-invariant
light-front Fock state wavefunctions (LFWFs) of the target
hadron~\cite{overlap2,overlap1}.  The longitudinal momentum transfer to the
target hadron is given by the ``skewness"
%parameter
variable $\zeta={Q^2\over 2 p\cdot q}.$
Since the incoming photon is space-like ($q^2 <0$)  and the final photon is
on-shell ($(q')^2=0$), the skewness is never zero in a physical experiment.
The DVCS process involves off-forward
hadronic matrix elements of light-front bilocal currents.  Accordingly,
in different kinematical regions, 
there is a diagonal parton-number conserving $n \to n$
overlap 
%in the kinematical regime $\zeta <x <1$ and $\zeta-1 < x < 0$ 
and an off-diagonal $n+1 \to n-1$ overlap 
%for $0<x<\zeta$,  
where the parton number is decreased by two.
Thus, given the LFWFs one then obtains a complete specification of all 
of the Generalized Parton Distributions (GPDs) measurable in DVCS, including 
their phase structure. The sum rules of DVCS, such as Ji's sum rule for 
angular momentum~\cite{Ji:1996ek} and the integral relations to electromagnetic 
and gravitational form factors are all explicitly satisfied in the  
light-front (LF)
formalism~\cite{overlap2,overlap1}.

In this paper we shall show how one can use measurements of the dependence
of the DVCS amplitude on the skewness
%parameter
variable $\zeta$ to obtain a novel
optical image of a hadron target, in  analogy to the way in which one
scatters optical waves  to form a diffraction pattern. 
In this context, it is useful to introduce a coordinate $b$ conjugate to the
momentum transfer $\Delta$ such that $ b \cdot \Delta = \frac{1}{2} b^+
\Delta^- + \frac{1}{2} b^- \Delta^+ - b_\perp \cdot \Delta_\perp$. Note that 
$ \frac{1}{2} b^- \Delta^+ = \frac{1}{2} b^- P^+ \zeta = \sigma \zeta $
where we have defined the boost invariant variable $\sigma$
which is an $`$impact parameter' in the longitudinal coordinate space. 
The Fourier transform
of  the  DVCS amplitude with respect to $\zeta$ allows one to determine the
longitudinal structure of the target hadron in terms of the variable
 $\sigma$.
% = {1\over 2} u^- P^+$, the boost-invariant longitudinal spatial 
%coordinate conjugate to the light-front longitudinal 
%momentum variable, $\zeta.$
 
Burkardt \cite{bur1,bur2} has studied the off-forward 
parton distribution function at zero longitudinal momentum transfer and fixed
longitudinal momentum fraction $x$ in the impact parameter ($b_\perp$)
space.  This relativistic  impact representation  on the light-front 
was introduced earlier by Soper \cite{soper} 
in the context of the Fourier Transform of the elastic form factor. 
We study the DVCS amplitude, which involves 
integration over $x$, in the
$\sigma$ space at fixed four-momentum transfer $ -t$.  Note that
experimentally DVCS amplitudes are measured as a function of $\zeta$ and
$-t$. Thus, our work is suited for the direct analysis of
experimental data and is  
complementary to the work of Burkardt and Soper. 
If one combines the longitudinal transform (at fixed $\Delta_\perp$) with
the Fourier
transform (FT) of the DVCS amplitude with respect to  $\Delta_\perp$
one can obtain a complete three-dimensional description of hadron
optics at fixed LF time. 

Recently, a 3D picture of the proton has been proposed in \cite{wigner} in a 
different approach, in terms of a Wigner distribution for the relativistic
 quarks and  gluons inside the proton. 
A major difference of this from our work is that the 
Wigner distributions are defined in the rest frame of the proton. 
%(which is not uniquely defined) 
Integrating out $k^-$ one gets the reduced Wigner
distributions which are not observable quantities in the quantum domain. 
Further integration over $k^\perp$ relates them to the FT of GPDs $H(x,\xi,t)$
 and $E(x,\xi,t)$, where 
$\xi =q^z/2E_q$ and $x$ is a special combination of off shell energy and 
momentum along $z$. On the other hand, 
we are taking the FT of the experimentally measurable 
DVCS amplitudes directly
and not the GPDs.

In principle, the LFWFs  of hadrons in QCD can be computed using
a nonperturbative method such as Discretized Light Cone Quantization (DLCQ)
where the LF Hamiltonian is diagonalized on a free Fock
basis~\cite{Brodsky:1997de}.  This has been accomplished for simpler
confining quantum field theories such as $QCD(1+1) $~ \cite{Hornbostel:1988fb}.
Models for the LFWFs of hadrons in $(3+1)$ dimensions displaying confinement
 at large
distances and conformal symmetry at short distances have been obtained using
the AdS/CFT method~\cite{Brodsky:2006uq}.
The light front longitudinal space  structure of
topological objects has been studied
in DLCQ \cite{topo}.

In order to illustrate  our general framework, we will present here an
explicit calculation of the  $\sigma$ transform of  virtual Compton scattering on the quantum fluctuations of a  lepton in QED at
one-loop order \cite{drell}, the same system which gives the Schwinger 
anomalous moment $\alpha/2 \pi$. This model has the advantage that it is 
Lorentz invariant, and thus it has
the correct relationship between the diagonal $n \to  n$ and the
off-diagonal $n - 1 \to  n +1 $ Fock state contributions to the 
DVCS amplitude. One can generalize this analysis
by assigning a mass $M$ to the external electrons and a different
mass $m$ to the internal electron lines and a mass $\lambda$ to the
internal photon lines with $M < m + \lambda$ for stability. 
In effect, we shall represent a spin-${1\over 2}$ system as a 
composite of a spin-${1\over 2}$
fermion and a spin-$1$ vector boson \cite{dis,dip,marc}.
We also will present numerical results for a composite hadron by taking a
derivative of the LFWFs with respect to the hadron's mass $M^2 .$
This simulates the behavior of a bound-state hadron by improving the fall-off
at the end points of the longitudinal momentum fraction $x$.
The summary of our main results will be given in this letter. A more
detailed analysis will be given in a forthcoming article \cite{forth}.

\vskip .2in
%%%%%%%%%%%%%%%%%%%%%%%%%%%%%%%%%%%%%%%%%%%%%%%%%%%%%%%%
\noindent {\bf DVCS in the  LF Formalism}
%%%%%%%%%%%%%%%%%%%%%%%%%%%%%%%%%%%%%%%%%%%%%%%%%%%%%%%%
\vskip .1in
The kinematics of the DVCS process has been given in detail in
\cite{overlap1,overlap2}.  One can work in a frame where the momenta
of the initial and final proton has a  $\Delta \to -\Delta$ symmetry
\cite{overlap1}. 
However, in this frame, the kinematics in terms of
the parton momenta becomes more complicated. Here, we choose the
frame  of Ref. \cite{overlap2}.

The virtual Compton amplitude $M^{\mu\nu}({\vec q_\perp},{\vec
\Delta_\perp},\zeta)$, {\it i.e.}, the transition matrix element of
the process $\gamma^*(q) + p(P) \to \gamma(q') + p(P')$, can be
defined from the light-cone time-ordered product of currents
\begin{eqnarray}
&&M^{\mu\nu}({\vec q_\perp},{\vec \Delta_\perp},\zeta)\ =\
i\int d^4y\,
e^{-iq\cdot y}\langle P'|TJ^\mu (y)J^\nu (0)|P\rangle \ ,
\label{com1j}
\end{eqnarray}
where the Lorentz indices $\mu$ and $\nu$ denote the polarizations of
the initial and final photons respectively. In the limit
$Q^2\to \infty$ at fixed $\zeta$ and $t$ the Compton amplitude is thus
given by
\begin{eqnarray}
\lefteqn{
M^{IJ}({\vec q_\perp},{\vec \Delta_\perp},\zeta)\ =\
\epsilon^I_\mu\, \epsilon^{*J}_\nu\,
M^{\mu\nu}({\vec q_\perp},{\vec \Delta_\perp},\zeta)\ =\
- e^2_{q}\ {1 \over 2\bar P^+}
\int_{\zeta-1}^1 dz\
}\nonumber
\\
&\times& \left\{ \ {t}^{IJ}(z,\zeta)\ {\bar U}(P')
\left[
H(z,\zeta,t)\ {\gamma^+}
 +
E(z,\zeta,t)\ {i\over 2M}\, {\sigma^{+\alpha}}(-\Delta_\alpha)
\right] U(P) \right\} , \label{com1a}
\end{eqnarray}
where $\bar P={1\over 2}(P'+P)$ and we take a frame in
which $q^+=0$. For DVCS, when
$Q^2$ is large compared to the masses and $-t$, we have,
\begin{equation}
{Q^2\over 2P \cdot q}=\zeta
\label{nn3}
\end{equation}
up to corrections in $1/Q^2$. Thus $\zeta$ plays the role of the
Bjorken variable in deeply virtual Compton scattering. For a fixed
value of $-t$, the allowed range of $\zeta$ is given by
\begin{equation}
0\ \le\ \zeta\ \le\
{(-t)\over 2M^2}\ \ \left( {\sqrt{1+{4M^2\over (-t)}}}\ -\ 1\ \right)\ .
\label{nn4}
\end{equation}
For simplicity we only consider one
quark with flavor $q$ and electric charge $e_{q}$. We here consider
the contribution of only the spin-independent GPDs $H$ and $E$.
Throughout our analysis we will use the $``$handbag" approximation where corrections to the hard quark propagator are neglected.

For circularly polarized initial and final photons ($I,\ J$ are
$\uparrow$ or $\downarrow)$) contributions only come from
\begin{eqnarray}
{t}^{\ \uparrow\uparrow}(z,\zeta)&=&
\phantom{-} \ {t}^{\ \downarrow\downarrow}(z,\zeta)\ =\
{1\over z-i\epsilon}\
+\ {1\over z-\zeta +i\epsilon}\ .
\label{com2p}
\end{eqnarray}
For a
longitudinally polarized initial photon, the Compton amplitude is of
order $1/Q$ and thus vanishes in the limit $Q^2\to \infty$. At order
$1/Q$ there are several corrections to the simple structure in
Eq.~(\ref{com1a}). We do not consider them here.

The generalized parton distributions
 $H$, $E$ are defined through matrix elements
of the bilinear vector and axial vector currents on the light-cone:
\begin{eqnarray}
\lefteqn{
F_{\lambda, \lambda'} (z,t)=  \int\frac{d y^-}{8\pi}\;e^{iz P^+y^-/2}\;
\langle P',\lambda' | \bar\psi(0)\,\gamma^+\,\psi(y)\,|P, \lambda \rangle
\Big|_{y^+=0, y_\perp=0} } \hspace{2em}
\label{spd-def} \nonumber\\
&=&
{1\over 2\bar P^+}\ {\bar U}(P', \lambda') \left[ \,
H(z,\zeta,t)\ {\gamma^+}
 +
E(z,\zeta,t)\
{i\over 2M}\, {\sigma^{+\alpha}}(-\Delta_\alpha)
\right]  U(P, \lambda)\ .
\label{defhe}
\end{eqnarray}
The off-forward matrix elements given by Eq. (\ref{defhe}) can be
expressed in terms of overlaps of LFWFs of the state
\cite{overlap1,overlap2}. For this, we take the state to be an electron in
QED at one loop and consider the LFWFs for this system.

The light-front Fock state wavefunctions corresponding to the
quantum fluctuations of a physical electron can be systematically
evaluated in QED perturbation theory. The state is expanded in
Fock space and there
are contributions from $\ket{e^- \gamma}$ and $\ket{e^- e^- e^+}$,
in addition to renormalizing the one-electron state. The
two-particle state is expanded as,
\begin{eqnarray}
\lefteqn{
\left|\Psi^{\uparrow}_{\rm two \ particle}(P^+, \vec P_\perp = \vec
0_\perp)\right> =
\int\frac{{\mathrm d} x \, {\mathrm d}^2
           {\vec k}_{\perp} }{\sqrt{x(1-x)}\, 16 \pi^3}
}
\label{vsn1}\\
&&
\left[ \ \ \,
\psi^{\uparrow}_{+\frac{1}{2}\, +1}(x,{\vec k}_{\perp})\,
\left| +\frac{1}{2}\, +1\, ;\,\, xP^+\, ,\,\, {\vec k}_{\perp}\right>
+\psi^{\uparrow}_{+\frac{1}{2}\, -1}(x,{\vec k}_{\perp})\,
\left| +\frac{1}{2}\, -1\, ;\,\, xP^+\, ,\,\, {\vec k}_{\perp}\right>
\right.
\nonumber\\
&&\left. {}
+\psi^{\uparrow}_{-\frac{1}{2}\, +1} (x,{\vec k}_{\perp})\,
\left| -\frac{1}{2}\, +1\, ;\,\, xP^+\, ,\,\, {\vec k}_{\perp}\right>
+\psi^{\uparrow}_{-\frac{1}{2}\, -1} (x,{\vec k}_{\perp})\,
\left| -\frac{1}{2}\, -1\, ;\,\, xP^+\, ,\,\, {\vec k}_{\perp}\right>\
\right] \ ,
\nonumber
\end{eqnarray}
where the two-particle states $|s_{\rm f}^z, s_{\rm b}^z; \ x, {\vec
k}_{\perp} \rangle$ are normalized as in \cite{overlap2}. The variables 
$s_{\rm f}^z$ and $s_{\rm b}^z$ denote the projection of the spins of the
constituent fermion and boson along the quantization axis, 
 and the variables $x$
and ${\vec k}_{\perp}$ refer to the momentum of the fermion. The
light cone momentum fractions $x_i= {k_i^+\over P^+}$  satisfy $0<
x_i \le 1$, $\sum_i x_i =1$. We employ the light-cone gauge $A^+=0$,
so that the gauge boson polarizations are physical. The
three-particle state has a similar expansion. Both the two- and
three-particle Fock state components are given in \cite{overlap2}.
We list here the two-particle wavefunctions for the spin-up electron
\cite{orbit,drell,overlap2}
\begin{equation}
\left
\{ \begin{array}{l}
\psi^{\uparrow}_{+\frac{1}{2}\, +1} (x,{\vec k}_{\perp})=-{\sqrt{2}}
\ \frac{-k^1+{i} k^2}{x(1-x)}\,
\varphi \ ,\\
\psi^{\uparrow}_{+\frac{1}{2}\, -1} (x,{\vec k}_{\perp})=-{\sqrt{2}}
\ \frac{k^1+{i} k^2}{1-x }\,
\varphi \ ,\\
\psi^{\uparrow}_{-\frac{1}{2}\, +1} (x,{\vec k}_{\perp})=-{\sqrt{2}}
\ (M-{m\over x})\,
\varphi \ ,\\
\psi^{\uparrow}_{-\frac{1}{2}\, -1} (x,{\vec k}_{\perp})=0\ ,
\end{array}
\right.
\label{vsn2}
\end{equation}
\begin{equation}
\varphi (x,{\vec k}_{\perp}) = \frac{e}{\sqrt{1-x}}\
\frac{1}{M^2-{{\vec k}_{\perp}^2+m^2 \over x}
-{{\vec k}_{\perp}^2+\lambda^2 \over 1-x}}\ .
\label{wfdenom}
\end{equation}
Similarly, the wavefunction for an electron with negative helicity
can also be obtained.

%At $x=1$, there
%are contributions from the overlap of one particle states which
%depend on the wavefunction renormalization constant $Z$. We have
%imposed  a cutoff on $x$ near this point.
%Also, in order to regulate the ultraviolet
%divergences, one has to introduce a regulator. Here, we use a cutoff
%$\Lambda$ on the transverse momentum $k^\perp$ as a regulator.

In the domain $\zeta <z <1$, there are diagonal $2 \to 2$ overlap
contributions to Eq. (\ref{defhe}), both helicity flip, $F^{22}_{+-}$
($\lambda' \ne \lambda$)
and helicity non-flip, $F^{22}_{++}$ ($\lambda'=\lambda$) \cite{overlap2}.
The GPDs $H_{(2\to 2)}(z,\zeta,t)$ and $E_{(2\to
2)}(z,\zeta,t)$ are zero in the domain $\zeta-1 < z < 0$, which
corresponds to emission and reabsorption of an $e^+$ from a physical
electron. Contributions to $H_{(n\to n)}(z,\zeta,t)$ and $E_{(n\to
n)}(z,\zeta,t)$ in that domain only appear beyond one-loop level.
This is because in the DVCS amplitude we have integrations over $z,~y^-,$ 
and $x$. When integration
over $y^-$ is performed, the fermion part of the bilocal current yields a
factor $\delta(z-x)$ and the anti-fermion part of the bilocal current yields
a factor $\delta(z+x)$. The latter contribution is absent in the one loop DVCS
amplitude of a electron target, which we consider in the present work.

The matrix elements $F^{22}_{++}$ and $F^{22}_{+-}$ are calculated
using the two-particle LFWFs given in Eq. (\ref{vsn2}).
The contributions in the domain, $0 < z < \zeta$, namely, $F^{31}_{+-}$ 
and $F^{31}_{++}$ come
from  overlaps of three-particle and one-particle LFWFs \cite{overlap2}.
%$Z$ is taken to be $1$ here. 
These are calculated using the three-particle
wavefunction. Explicit
expressions of all the above matrix elements will be given in \cite{forth}.

We calculate the DVCS amplitude given by Eq. (\ref{com1a}) using the
off-forward matrix elements calculated above. 
In order to regulate the ultraviolet
divergences, we use a cutoff
$\Lambda$ on the transverse momentum $k^\perp$.
The real and imaginary parts
are calculated separately using the prescription
\be
\int_0^1 dx {1\over x-\zeta+i \epsilon}F(x,\zeta)= P\int_0^1 dx 
{1\over x-\zeta} F(x,\zeta) - i \pi F(\zeta,\zeta).
\ee
Here $P$ denotes the principal value defined as
\be
P \int_0^1 dx {1\over x-\zeta}F(x,\zeta)=\lim_{\epsilon \to 0}\Big[
\int_0^{\zeta-\epsilon}{1\over x-\zeta} F(x,\zeta)+\int_{\zeta+\epsilon}^1
{1\over x-\zeta} F(x,\zeta)\Big]
\ee 
where 
\be 
F(x,\zeta)&=&F^{31}_{ij}(x,\zeta,\Delta_\perp),~{\rm for}~ 0<x<\zeta \nonumber\\
         &=& F^{22}_{ij}(x,\zeta,\Delta_\perp),~{\rm for}~ \zeta<x<1\nonumber
\ee 
 with $ij=++$ for helicity non-flip and $ij=+-$ for helicity flip amplitudes. Since the off-forward matrix elements are continuous at $x=\zeta$,
  $F(\zeta,\zeta)=F^{22}_{ij}(x=\zeta,\zeta,\Delta_\perp)=F^{31}_{ij}(x=\zeta,\zeta,\Delta_\perp)$.
Note that
the principal value prescription cannot be used at $x=0$. We take a small 
cutoff at this point for the numerical calculation. The 
off-forward matrix elements $F^{31}$ (which contribute in the kinematical 
region $0 < x < \zeta$) vanish as $x \rightarrow 0$, as a result there is no
logarithmic divergence at this point for nonzero $\zeta$. But, we need to 
be careful here as
when we consider the Fourier transform in $\sigma $ space, $\zeta$ can go to zero and divergences from small $x$ can occur from $F^{22}$ which is finite and nonzero
at $x,\zeta \to 0$. 

The imaginary part of the amplitude when the electron helicity is
not flipped is then given by 
\be {\mathrm{Im}} [M_{++}] (\zeta,
\Delta_\perp) =
   \pi e^2 F^{22}_{++}(x=\zeta,\zeta,\Delta_\perp). 
\ee 
A similar expression holds in the case
when the electron helicity is flipped ($ {\mathrm{Im}} [M_{+-}]
(\zeta, \Delta_\perp) $) in which $ F_{++}$ are replaced by $F_{+-}$.
%The last line above follows from the fact that the off-forward matrix 
%elements are continuous at $x=\zeta$. 
The helicity-flip DVCS amplitude is 
proportional to $(\Delta_1-i
\Delta_2)$ \cite{forth}.  Without any loss of generality, the plots
 for these amplitudes are
presented with $\Delta_2 =0$.  The imaginary part
receives contributions at $x=\zeta$. The other regions of $x$
contribute to the real part. It is to be emphasized that we are
using the handbag approximation of the DVCS amplitude. Contributions
from the Wilson lines are in general not zero, and they can give
rise to new phase structures as seen in single-spin
asymmetries \cite{Brodsky:2002cx}.

The real  part of the DVCS amplitude in our model is given by
\be
{\mathrm{Re}}~
[M_{++}]~ (\zeta, \Delta_\perp) &=& -e^2  \int_{\eps}^{\zeta-\eps_1} ~dx~
F^{31}_{++} (x,\zeta, \Delta_\perp) ~\Big [~ {1\over x}~
 +  ~{1\over{x-\zeta}}~  \Big ] \nonumber\\
&&- e^2
\int_{\zeta+\eps_1}^{1-\eps}~dx~ F^{22}_{++} (x,\zeta, \Delta_\perp) \Big [
~{1\over x}~ + ~{1\over{x-\zeta}}~  \Big ].
\ee
A similar expression holds for the helicity flip DVCS amplitude.
The cutoff dependence at $x=\zeta$ in the principal
value prescription gets canceled explicitly and, as a result,
the DVCS amplitude is independent of the cutoff  $\eps_1$.

%%%%%%%%%%%%%%%%%%%%%%%%%%%%%%%%%%%%%%%%%%%%%%%%%%%%%
\vskip .2in
\noindent{\bf Calculation of the $\sigma$ Fourier Transform}
\vskip .1in

In order to obtain the DVCS amplitude in longitudinal coordinate 
space, we take a
Fourier transform in $\zeta$ as,
\be
A_{++} (\sigma, t) =
{1\over 2 \pi } \int_{\eps_2}^{1-\eps_2} d\zeta ~ e^{{i}
\sigma \zeta }~ M_{++}
(\zeta, \Delta_\perp),\nonumber\\
A_{+-} (\sigma, t) = {1\over 2 \pi }
\int_{\eps_2}^{1-\eps_2} d\zeta ~ e^{{i} \sigma \zeta }~ M_{+-}
(\zeta, \Delta_\perp),
\ee
where $\sigma = {1\over 2} P^+ b^-$ is the (boost invariant) longitudinal distance on
the light-cone and the FTs are performed at a fixed invariant momentum 
transfer squared $-t$.
We have imposed cutoffs at $\eps=\eps_1=\eps_2/2=0.001$
for the numerical calculation. 

A detailed discussion of the
cutoff scheme will be given in \cite{forth}. 
%In the frame we work, $\sigma$ is the longitudinal distance between the 
%initial and final proton (target). The magnitude of the resulting boost-
%invariant amplitude as a function of this distance is the measure of the 
%ability of the hadron wavefunction to both emit and absorb the quark with 
%the physical properties matched to the initial and final photon kinematics.

All Fourier transforms (FT) have been performed by numerically calculating the
Fourier sine and cosine transforms and then calculating the resultant by
squaring them, adding and taking the square root, thereby yielding the
Fourier Spectrum (FS).
In Fig. 1, we have shown the FS of the imaginary part of the DVCS
amplitude 
%with respect to $\zeta$ 
for $M = 0.51$ MeV, $m = 0.5$ MeV
and $\lambda = 0.02$ MeV. (a) is the helicity non-flip and (b) is the helicity
flip part of the amplitude. We have divided the amplitude by
the normalization constant $e^4/ (16\pi^3)$ and have taken $\Lambda=Q=10$ MeV.
The helicity non-flip amplitude  depends on the scale $\Lambda$
logarithmically and  the scale dependence is suppressed in the helicity-flip
part. 
As seen in Fig. 1(a), the FS of the imaginary part of the helicity non-flip
amplitude displays a diffraction pattern in $\sigma$. The peak initially
increases with increasing $-t$, but then decreases as $-t$ increases
further. The latter behavior warrants further study. In contrast, we see in
Fig. 1(b) that there is no diffraction pattern in the FS of the imaginary
part of the helicity flip amplitude. This is due to different behavior with
respect to $\zeta$ of the respective amplitudes \cite{forth}. Further,
note that in this
case the peak monotonically increases with increasing $-t$. One reason for
this may be the presence of the extra factor of $\Delta_\perp$ in the
helicity flip amplitude compared to the helicity non-flip amplitude.
\begin{figure}[t]
%\begin{minipage}[c]{0.9\textwidth}
\includegraphics[width=7cm,height=7cm,clip]{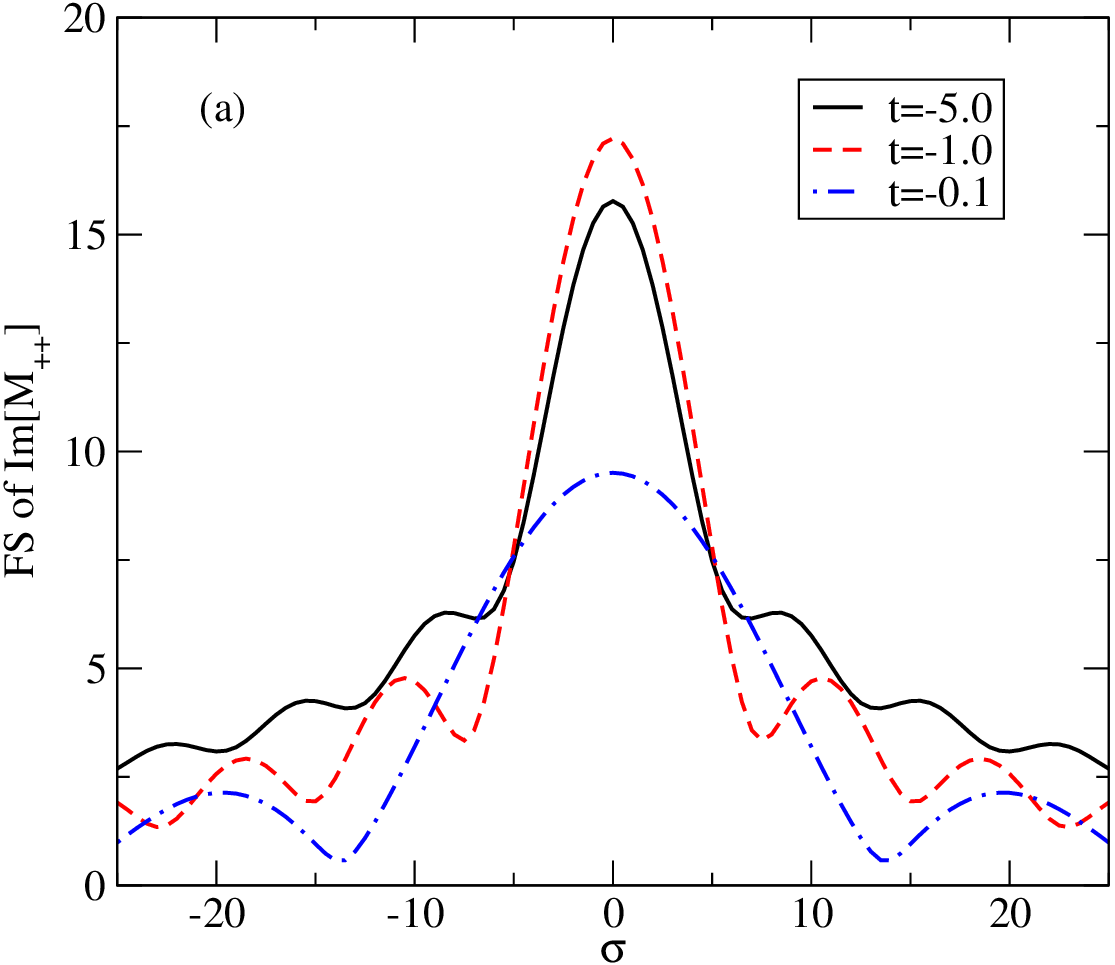}%
\hspace{0.2cm}%
\includegraphics[width=7cm,height=7cm,clip]{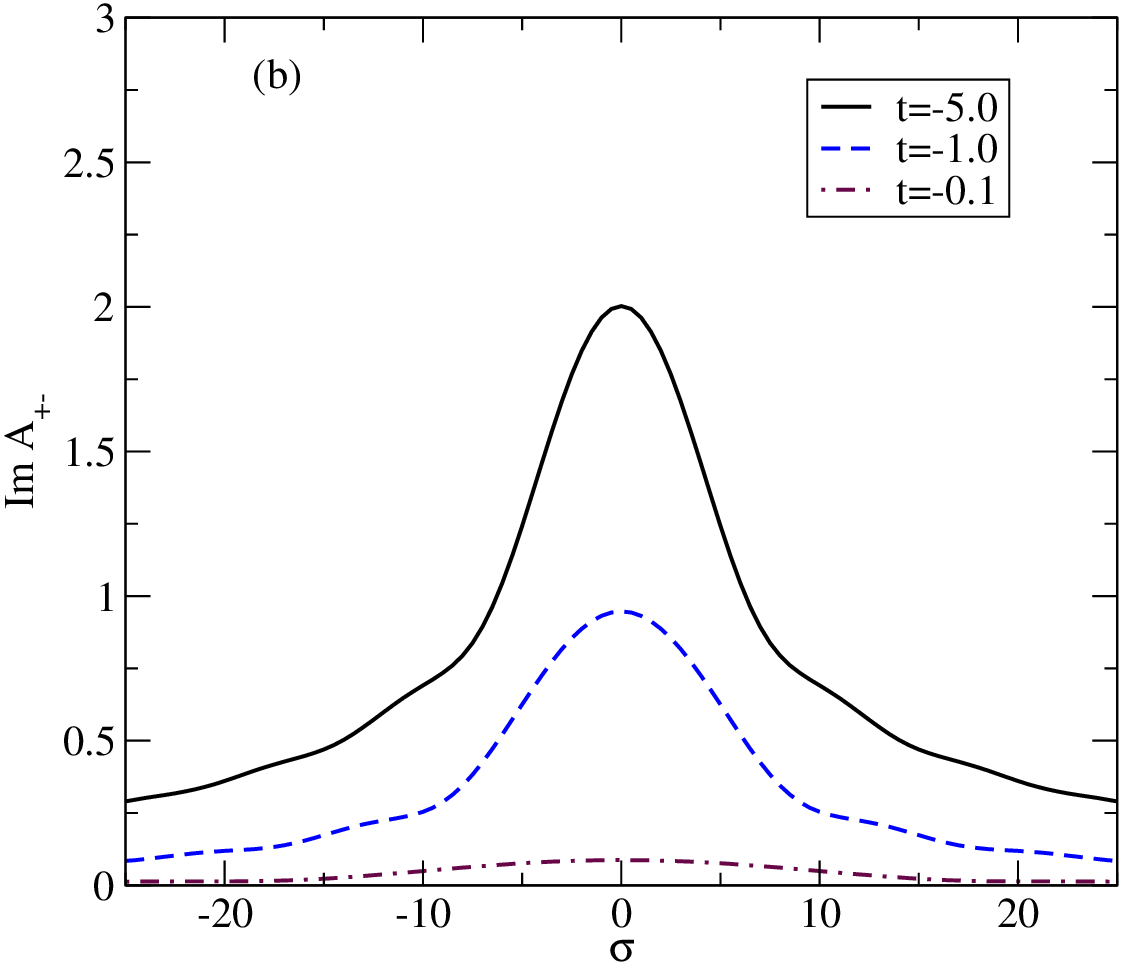}
\caption{\label{fig1} Fourier spectrum of the imaginary part of
the DVCS amplitude of an electron vs. $\sigma$
for $M=0.51$ MeV, $m=0.5$ MeV,
$\lambda=0.02$  MeV, (a) when the electron helicity is not flipped;
(b) when the helicity is flipped.
The parameter $t$ is in ${\mathrm{MeV}^2}$.}
\end{figure}

In Fig. 2 we have plotted the Fourier Spectrum  of the real part of the
DVCS amplitude vs. $\sigma$  for $M = 0.51$ MeV, $m = 0.5$ MeV
and $\lambda = 0.02$ MeV.
% As mentioned above, the
%helicity non-flip part of the amplitude depends on $\Lambda$ logarithmically.
%We have used a fixed value of $\Lambda=Q=10$ MeV.
For the same $\mid t \mid$,
the behavior is
independent of $Q$ when $\mid t \mid < m^2$. For each $Q$, the peak at
$\sigma=0$ is sharper and higher as $\mid t \mid $ increases, the
number of minima within the same $\sigma$ range also increases. As seen
in Fig. 2, the FS of both
the helicity flip and helicity non-flip parts show a diffraction pattern.
%this is because unlike the imaginary part, the real part of the
%helicity flip amplitude does not vanish at $\zeta=0$ \cite{forth}.
%The FS of the real part of the helicity flip amplitude in $\sigma$ also
%decreases with $t$ for higher $t$.
\begin{figure}[t]
\centering
\includegraphics[width=7cm,height=7cm,clip]{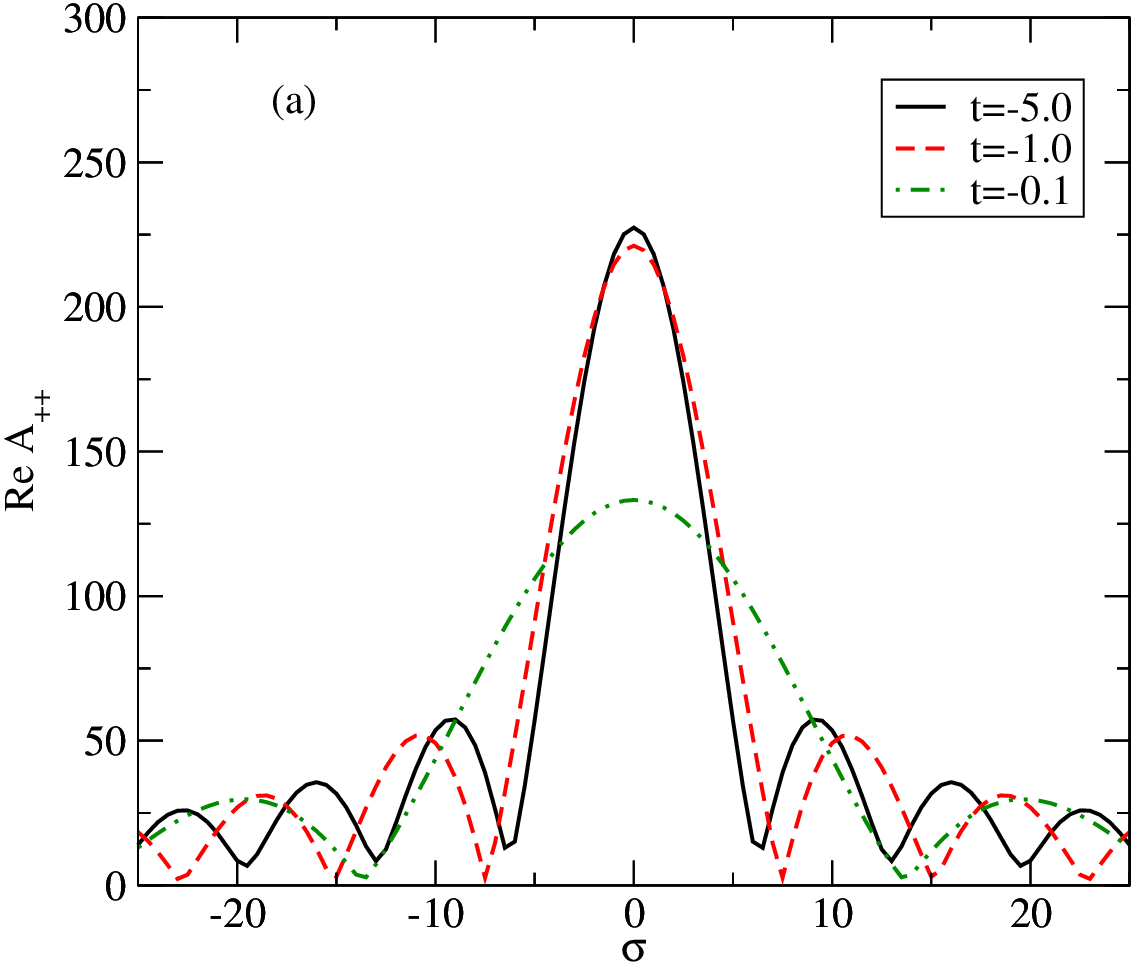}%
\hspace{0.2cm}%
\includegraphics[width=7cm,height=7cm,clip]{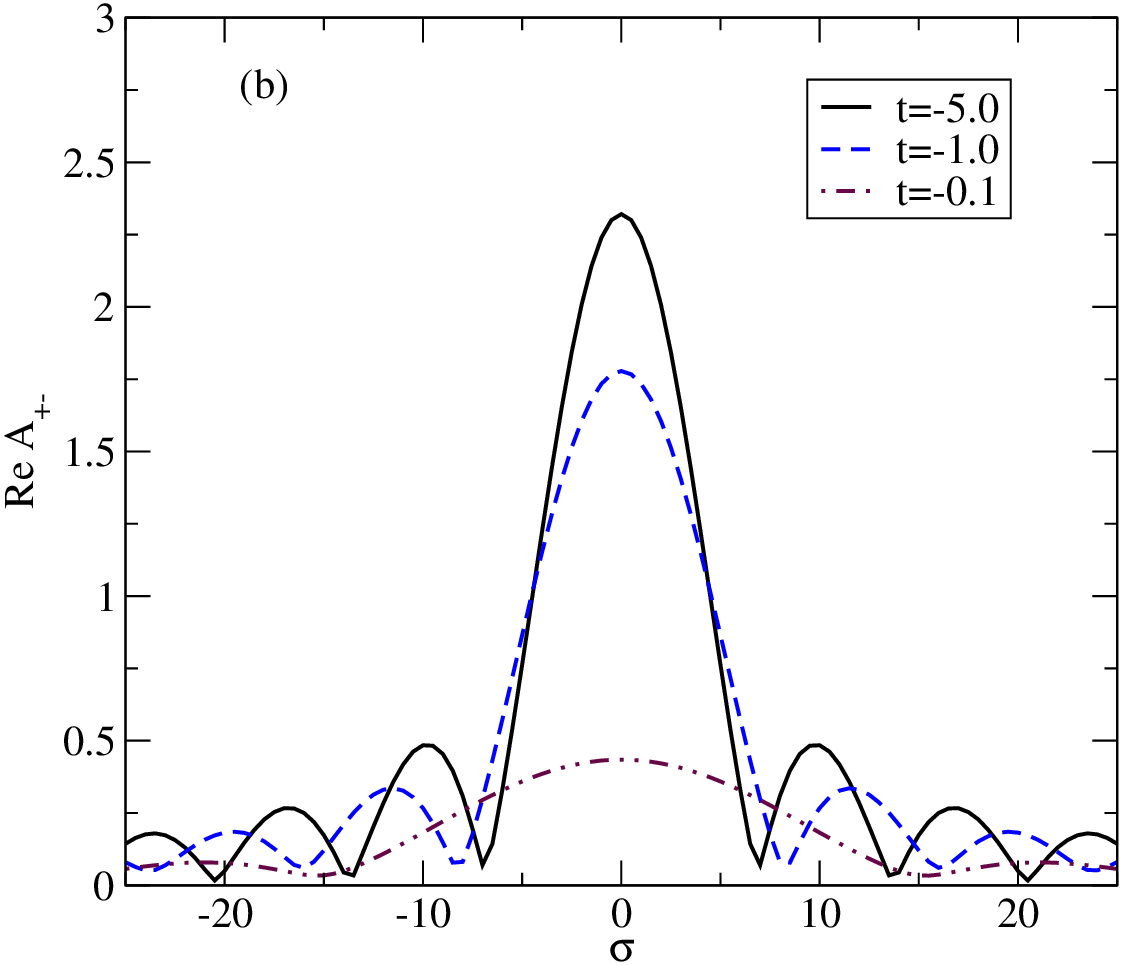}
%\end{minipage}%
\caption{\label{fig2} Fourier spectrum of the real part of
the DVCS amplitude of an electron vs. $\sigma$
for $M=0.51$ MeV, $m=0.5$ MeV,
$\lambda=0.02$  MeV, (a) when the electron helicity is  not flipped;
(b) when the helicity is flipped.
The parameter $t$ is in ${\mathrm{MeV}^2}$.}
\end{figure}

\vspace{.2cm}
The DVCS amplitude for an electron-like state at one loop
has potential singularities at $x=1$. As
mentioned above, we have used cutoffs at $x=0,1$. The cutoff at
$x=0$ is imposed for the numerical integration.
In the $2-$ and $3$-body LFWFs, the bound-state mass squared $M^2$ appears
in the denominator.  Differentiation of the LFWFs with respect to $M^2$
increases the fall-off of the wavefunctions near the end points $x=0,1$
and mimics the hadronic wavefunctions. In this way, the cutoff dependency is
removed. Differentiating once with respect
to $M^2$ simulates a meson-like wavefunction and another differentiation
simulates a proton wavefunction. Convolution of these wavefunctions
in the same way as we have done for the dressed electron wavefunctions
will simulate the corresponding DVCS amplitudes for bound state hadrons.
One has to note that differentiation of the single particle wave function
yields zero and thus there is no $3-1$ overlap contribution to the DVCS
amplitude in this hadron model. It is to be noted that in recent 
holographic models from AdS/CFT as well \cite{Brodsky:2006uq}, only 
valence LFWFs are constructed.  
  
The equivalent but easier way is to differentiate the
DVCS amplitude with respect to the initial and final state masses.
Here we calculate the quantity $M_F^2 {\partial \over \partial M_F^2}
M_I^2 {\partial \over \partial M_I^2} A_{ij} (M_I, M_F)$ where $M_I, M_F$ are
the initial and final bound state masses. For numerical computation, we use
the discrete (in the sense that the denominator is small and finite but not 
limiting to zero) version of the differentiation
\be
M^2 {\partial A\over \partial M^2} = {\bar M}^2 {A(M_1^2)-A(M_2^2)\over
\delta M^2}
\ee
where ${\bar M}^2={(M_1^2+M_2^2)\over 2}$ and $\delta M^2 = (M_1^2-M_2^2)$.
We have taken $M_{I1}, M_{F1}=150+1$, $M_{I2}, M_{F2}=150-1$ MeV
and fixed parameters $M=150$ and $m=\lambda=300$ MeV.
In Figs. 3 and 4 we have shown the DVCS amplitude of the simulated hadron
model, both as a function of $\zeta$ and after taking  the FT in $\zeta$. In
Fig. 4 (c), we have plotted the structure function $F_2(x)$ in this model.
The wave function is normalized to $1$.
Recall that the
$\gamma^* p \to \gamma p$ DVCS amplitude has both real \cite{real}
and imaginary parts \cite{imag}. The imaginary part requires a 
non-vanishing LFWF at $x'={x-\zeta\over 1-\zeta} = 0$. If we consider a
dressed electron, the imaginary part from the pole at $x = \zeta $ survives
because of the numerator ${1\over x-\zeta}$ factor
in the electron's LFWF. This numerator behavior reflects the
spin-1 nature of the constituent boson. The $x -\zeta \to 0$ singularity
is shielded when we differentiate the final state  LFWFs with respect to
$M^2$ and, as a result,  the imaginary part of the amplitude vanishes 
in this model. We thus have constructed a model where the DVCS amplitude is
purely real.   
It is interesting that the forward virtual Compton amplitude $\gamma^* p
\to \gamma^* p$ (whose imaginary part gives the structure function) does not
have this property.  The pole at $x = \zeta$ is not shielded since the
initial and final $n=2$ LFWFs are functions of $x$.  If instead we consider the
differentiation with respect to the internal fermion mass $m^2$ rather than
the bound state mass $M^2$, although it does not improve the wavefunction
behavior at the endpoint $x=0$, we can generate a
model with both real and imaginary parts of the DVCS amplitudes. 
It is worthwhile to point out that in general the LFWFs for a hadron 
may be non-vanishing at the end points \cite{dalley}, and recent measurements 
of single spin asymmetries suggest that the GPDs are non-vanishing at 
$x= \zeta$ \cite{exp}. A more 
realistic estimate would require non-valence Fock states \cite{crji}.
\begin{figure}
\centering
\includegraphics[width=7cm,height=7cm,clip]{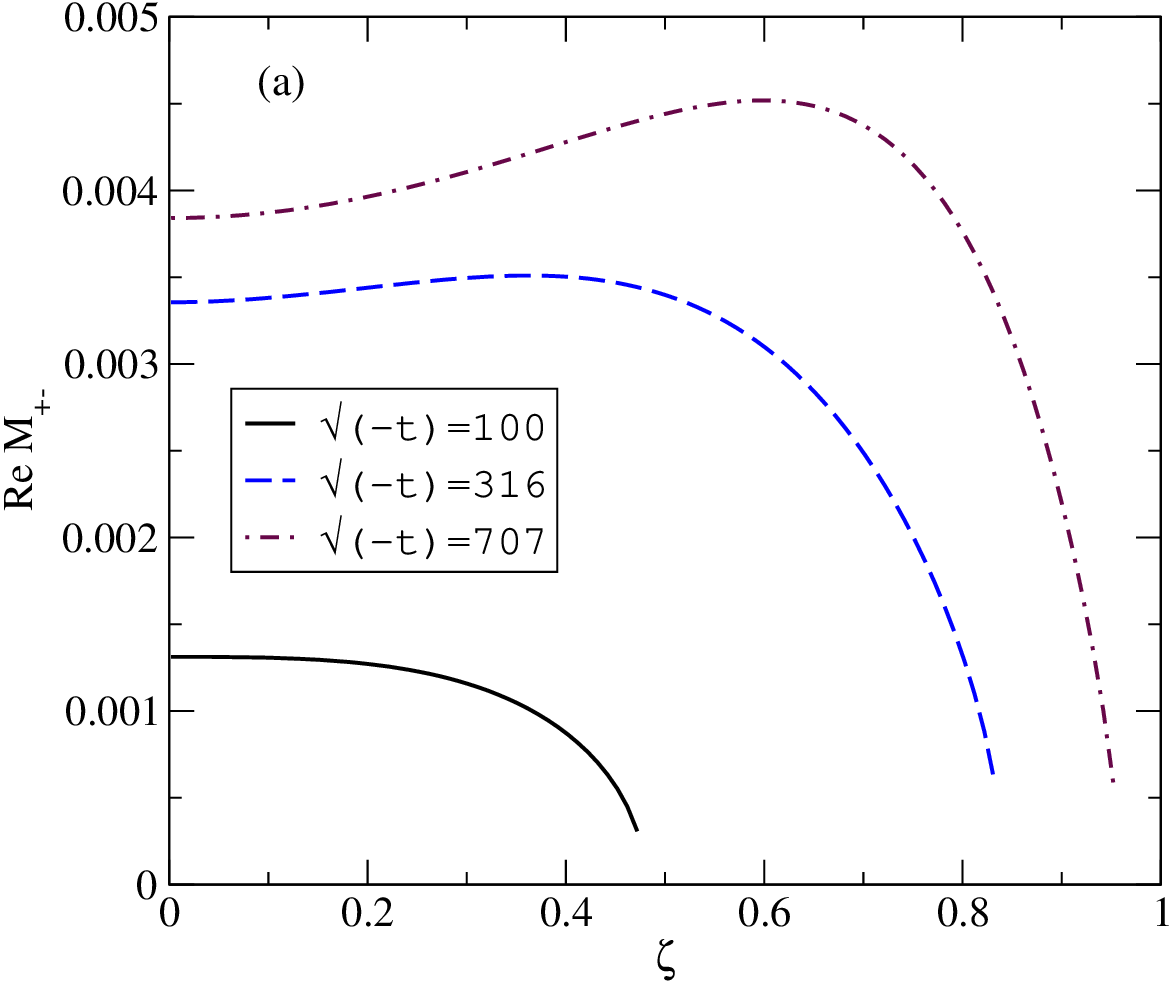}%
\hspace{0.2cm}%
\includegraphics[width=7cm,height=7cm,clip]{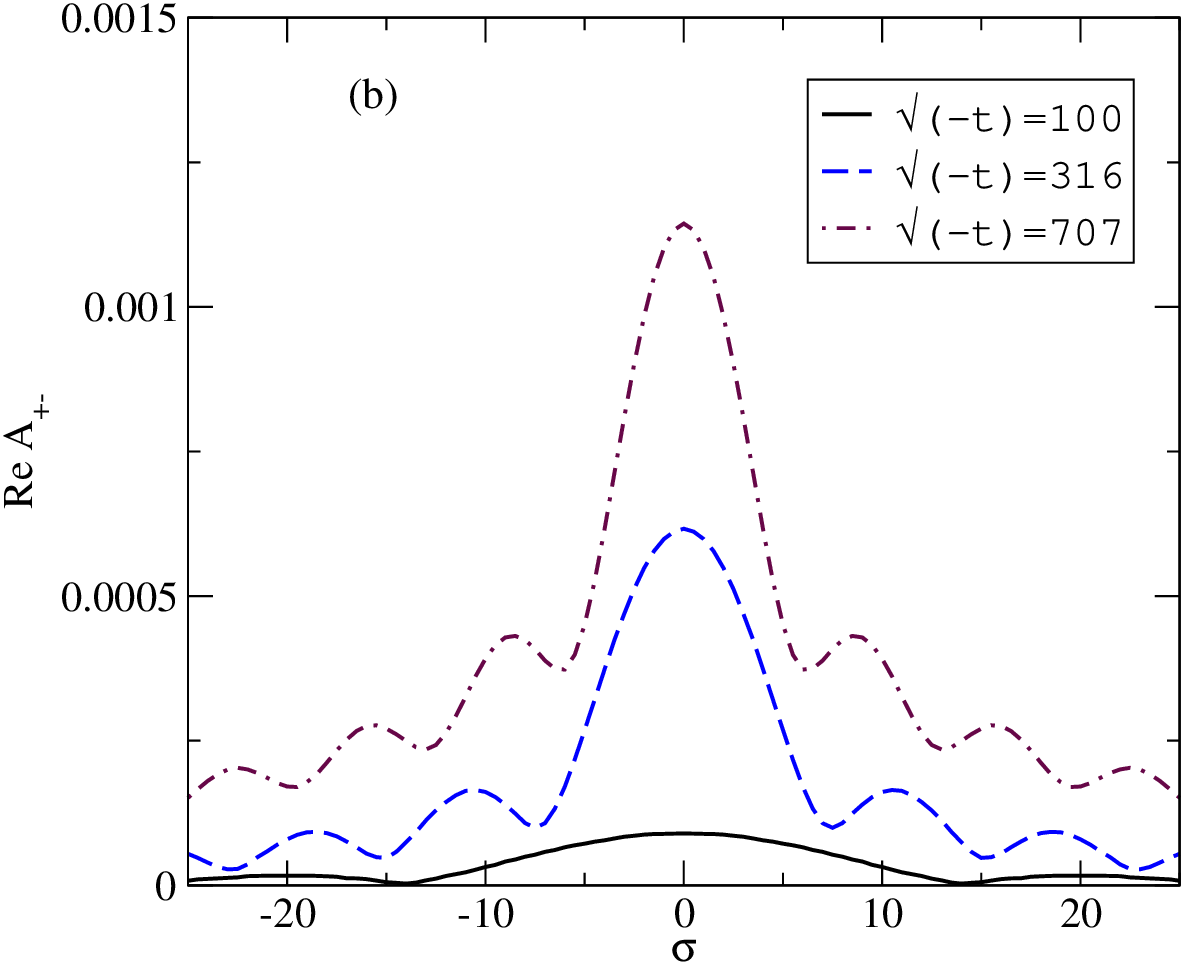}
%\end{minipage}%
\caption{\label{fig3} Real part of the DVCS amplitude for the
simulated meson-like bound state.
The parameters are $M=150, m=\lambda=300$ MeV.
(a) Helicity flip amplitude  vs. $\zeta$, (b)
Fourier spectrum of the same vs. $\sigma$.
The parameter $t$ is in ${\mathrm{MeV}^2}$.}
\end{figure}
\begin{figure}[t]
\centering
\includegraphics[width=7cm,height=7cm,clip]{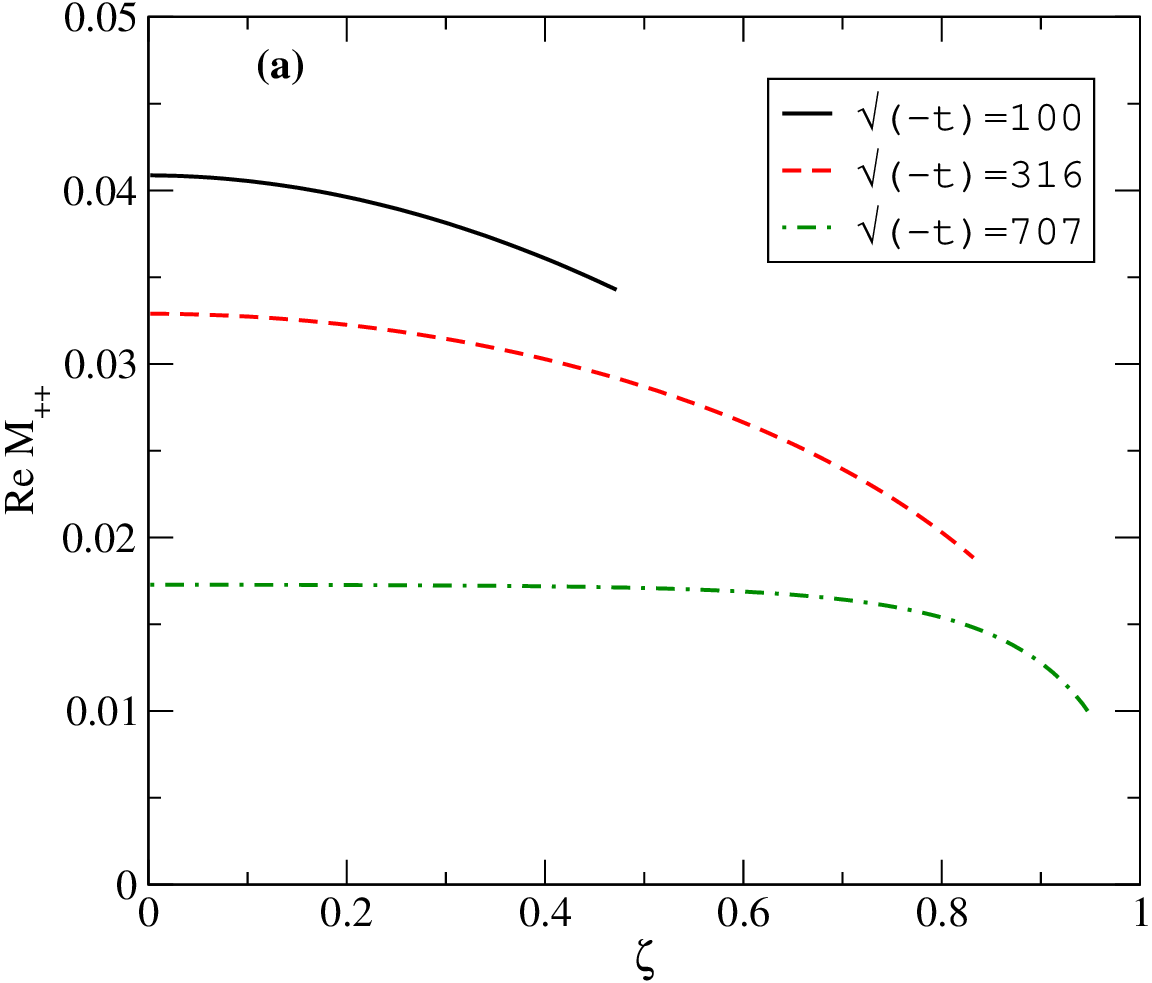}%
\hspace{0.2cm}%
\includegraphics[width=7cm,height=7cm,clip]{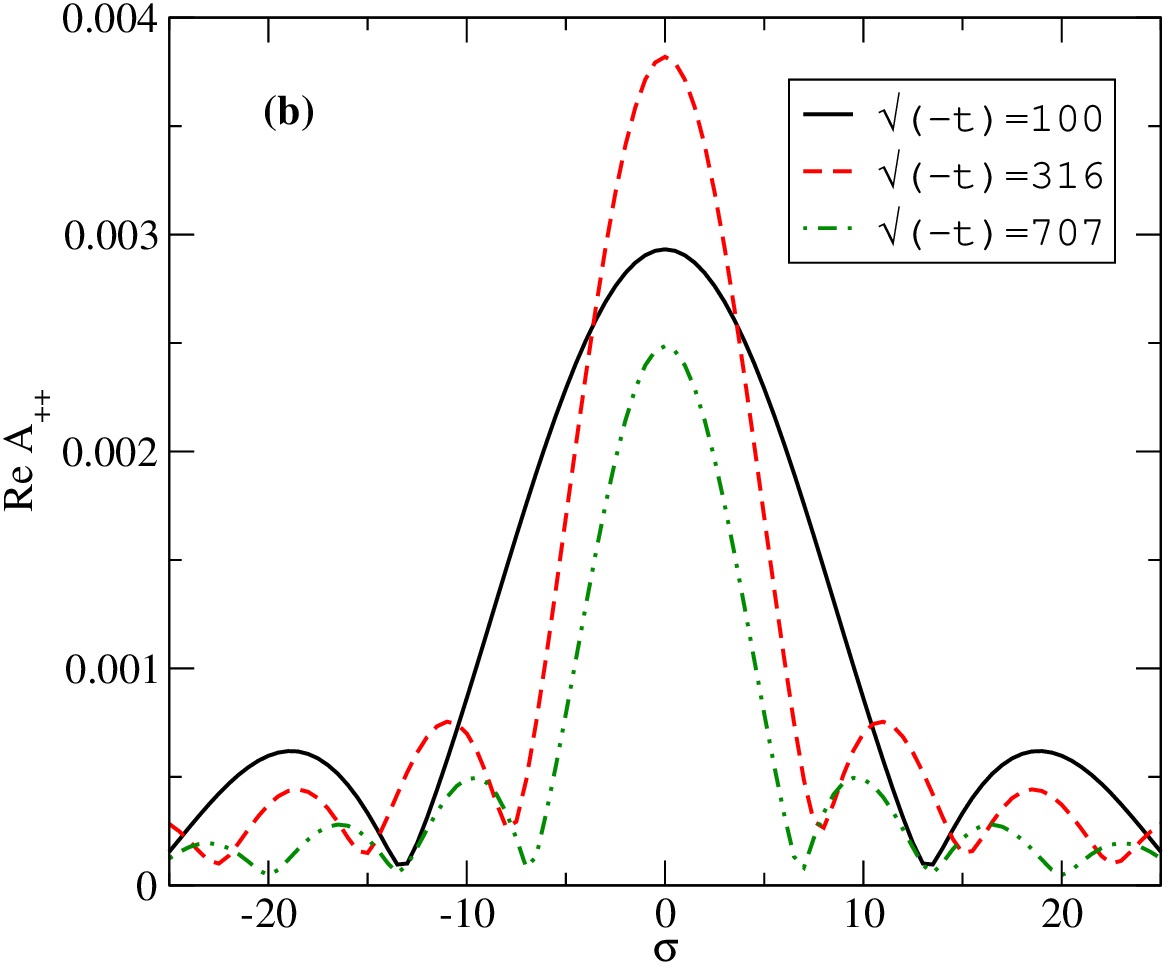}
\begin{minipage}[c]{0.5\textwidth}
%\centering
\vspace{0.2cm}%
\includegraphics[width=7cm,height=7cm,clip]{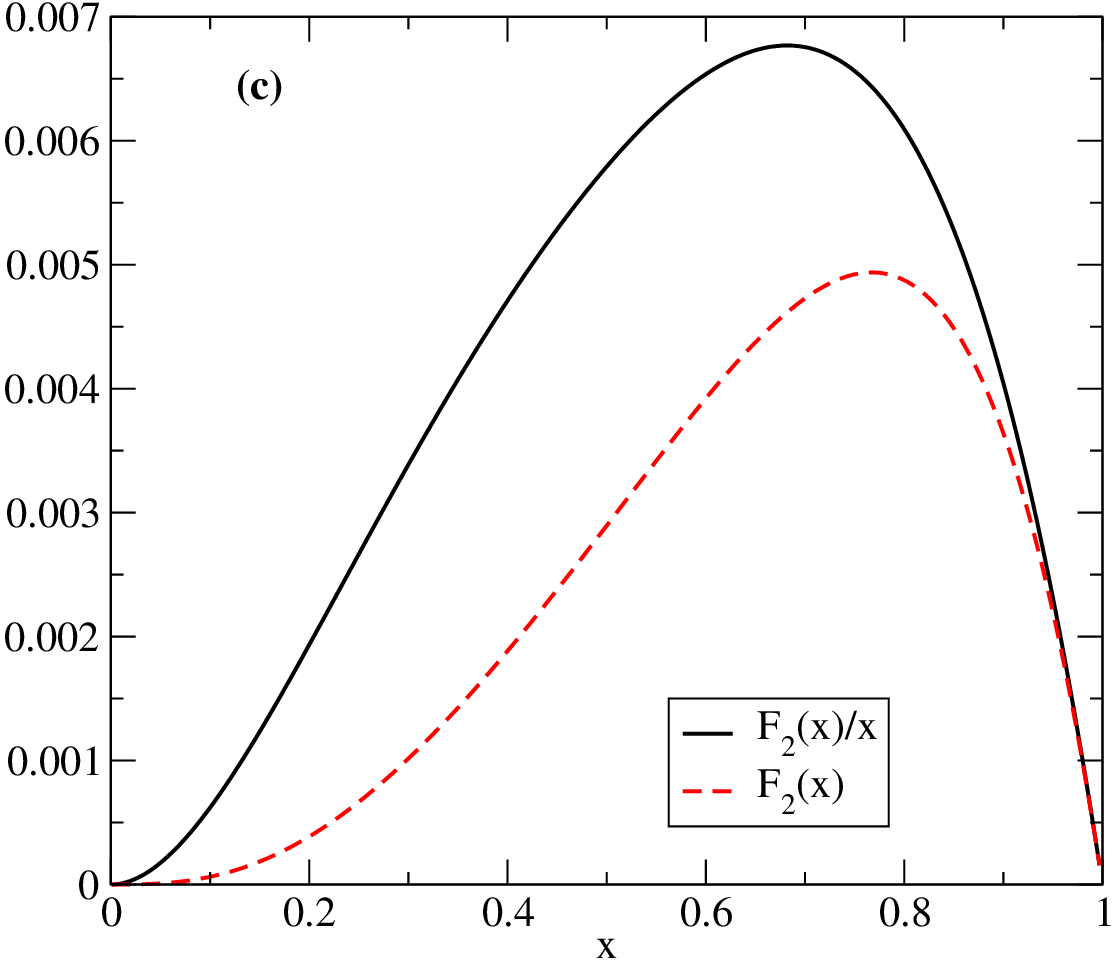}
\end{minipage}%
\caption{\label{fig4} Real part of the DVCS amplitude for the simulated
meson-like bound state.
The parameters are $M=150, m=\lambda=300$ MeV.
(a) Helicity non-flip amplitude  vs. $\zeta$, (b)
Fourier spectrum of the same vs. $\sigma$, (c) Structure function vs. x.
The parameter $t$ is in ${\mathrm{MeV}^2}$.}
\end{figure}
%\vspace{.2cm}

From the plots, we propose an optics analog of the behavior of the
Fourier Spectrum of the DVCS amplitude.
In fact, the similarity of  paraxial optics and
quantum fields on the light cone  was first explored long ago  in
\cite{sudarshan1,sudarshan2}. 
%In the case of DVCS, we are
%effectively looking at the interference pattern between the plane
%wave of the initial virtual photon and the plane wave of the
%outgoing real photon. 
In the case of DVCS, the final-state proton wavefunction is
modified relative to the initial state  proton wavefunction because of the
momentum transferred to the quark in the hard Compton scattering.
The quark momentum undergoes changes in both longitudinal ($\zeta P^+$) and
transverse ($\Delta_\perp$) directions. We remind the reader that, to keep
close contact with experimental analysis, we have kept $-t$ fixed while
performing the Fourier transform over the $ \zeta$ variable. 
%The change in quark momentum along the longitudinal direction
%$\zeta$ can be Fourier transformed to a boost-invariant distribution
%in the longitudinal light-front coordinate  $\sigma = {1\over 2} 
%P^+b^-$.
% the quantum mechanical spatial coordinate conjugate to the
%light-cone momentum fraction $x = {k^+\over P^+}.$ 
Note that the integrals over $x$ and $\zeta$ are of finite range. More
importantly the upper limit of $\zeta$ integral is $\zeta_{max}$ which in
turn is determined by the value of $-t$. The  finiteness of slit width
is a {\em necessary} condition for the occurrence of diffraction pattern in
optics. Thus when integration is performed over the range from $0$ to 
$\zeta_{max}$, this finite range acts as a slit of finite width and provides
a necessary condition for the occurrence of diffraction pattern in the
Fourier transform of the DVCS amplitude.
% The situation in DVCS amplitude is
%more complicated because of the presence of
%wave functions. 
%Nevertheless,
When a diffraction pattern is produced, in analogy with single slit
diffraction, we expect the position of the first minimum to be inversely
proportional to $\zeta_{max}$. Since $\zeta_{max}$ increases with $-t$, we
expect the position of the first minimum to move to a smaller value of
$\sigma$, in analogy with optical diffraction. In the
case of the Fourier Spectrum of DVCS on the quantum
fluctuations of a lepton target in QED, and also in the
corresponding hadronic model, one sees that the diffractive
patterns in $\sigma$ sharpen and the positions of the first minima
typically move in with increasing momentum transfer. Thus the
invariant longitudinal size of the parton distribution becomes
longer and the shape of the conjugate light-cone momentum
distribution becomes narrower with increasing $\mid t \mid$.
Regarding the diffraction patterns observed in the Fourier Spectrum 
of the DVCS amplitude, we further note that  for fixed $-t$, higher minima appear at
positions which are integral multiples of the lowest minimum. This 
further supports the  analogy with diffraction in optics.

We can study the diffraction pattern in $\sigma$ as a function of $t$
or $\Delta^2_\perp$  in order to register the effect of a change
in transverse momentum resulting from the Compton scattering.  If one
Fourier transforms in  $\zeta$ at fixed $\Delta_\perp$ and then Fourier
transforms the change in transverse momentum $\Delta_\perp$
to impact space $b_\perp$ \cite{bur1,bur2}, then one would have the
analog of a three-dimensional scattering center. In this sense,
scattering photons in DVCS provides the complete Lorentz-invariant
light front coordinate space structure  of a hadron.

This work was supported, in part, by the US Department of Energy grant Nos.
DE-AC02-76SF00515, DE-FG02-97ER-41029, and DE-FG02-87ER40371
and at the University of
California, Lawrence Livermore National Laboratory under contract No.
W-7405-Eng-48. This work was also supported in part by the Indo-US
Collaboration project jointly funded by the U.S. National Science
Foundation (NSF) (INT0137066) and the Department of Science and
Technology, India (DST/INT/US (NSF-RP075)/2001).
% JPV suggests replacing the following, for brevity,
%The work of DC was supported in part by US Department of Energy
%under grant No. DE-FG02-97ER-41029.
%with
%The work was also supported in part by US Department of Energy
% under grants No. DE-FG02-97ER-41029 and DE-FG02-87ER40371.

%%%%%%%%%%%%%%%%%%%%%%%%%%%%%%%%%%%%%%%%%%%%%%%%%%%%%%%%%%%%%%%%%%%%%%%%

%%%%%%%%%%%%%%%%%%%%%%%%%%%%%%%%%%%%%%%%%%%%%%%%%%%%%%%%%%%%%%%%%%%%%%%%%%%
\end{document}